\documentclass[11pt,a4paper]{article}
\usepackage{jcappub}
\usepackage[utf8]{inputenc}
\usepackage{url}

\title{Dark matter velocity dispersion effects on CMB and matter power spectra}

\author[a]{O.~F.~Piattella,}
\author[a]{L.~Casarini,}
\author[a]{J.~C.~Fabris,}
\author[b]{and J. A. de Freitas Pacheco}

\affiliation[a]{Department of Physics, Universidade Federal do Esp\'irito Santo,\\
avenida Ferrari 514, 29075-910 Vit\'oria, Esp\'irito Santo, Brazil}
\affiliation[b]{Universit\'e de Nice-Sophia Antipolis, Observatoire de la C\^ote d'Azur,\\ 
Laboratoire Lagrange, Nice Cedex 4, France}

\emailAdd{oliver.piattella@pq.cnpq.br}

\abstract{Effects of velocity dispersion of dark matter particles on the CMB TT power spectrum and on the matter linear power spectrum are investigated using a modified CAMB code. Cold dark matter originated from thermal equilibrium processes
does not produce appreciable effects but this is not the case if particles have a non-thermal origin. A cut-off in 
the matter power spectrum at small scales, similar to that produced by warm dark matter or that produced in the late 
forming dark matter scenario, appears as a consequence of velocity dispersion effects, which act as a pressure perturbation.}

\keywords{Dark Matter, Velocity Dispersion, CMB power spectrum, CAMB.}

\arxivnumber{1507.00982}

\begin{document}
\maketitle

\flushbottom


\section{Introduction}

Different cosmological and astrophysical observations point towards the existence of Dark Matter (DM), a non-electromagnetic interacting form of matter that is responsible for structure formation and other gravity-related phenomena. Among these, we may cite the observed high velocity dispersion in clusters of galaxies, the confinement of the hot gas detected in these objects and the flat rotation curve of spiral galaxies. In particular, the modelling of the rotation curve of the Milky Way requires, besides the known baryonic components, the presence of a massive dark halo \cite{Iocco:2015xga, Paulo2015}. 

The nature of DM is one of the greatest challenges in cosmology and physics today. Since standard model relics are not able to explain the required cosmic abundance, DM is possibly related to extensions or modifications of the canonical model or can even require new physics. The most promising particle candidates are WIMPs (Weakly Interacting Massive Particles), which are interesting because a stable relic with mass close to the electroweak scale ($\sim 100$ GeV) has an expected abundance today able to explain the observations (i.e. $\Omega_{\rm dm} \sim 0.3$). Unfortunately, despite the great efforts deployed by different experimental groups trying to detect signals from the interaction between these particles and those of the standard model, until now no positive result has been reported. For a review on DM particles, see for instance ref.~\cite{Bertone:2010zza}.

In order to trigger structure formation, DM must behave like a pressureless fluid or, in other words, to have a quite small velocity dispersion, reason why it is called cold DM (CDM). However, in the early universe WIMPs are still coupled to the primordial plasma, having thus a non-negligible pressure. This is consequence of interactions with standard model particles (essentially leptons) that produce damping effects in the primordial density fluctuation power spectrum on scales smaller than a damping (or diffusion) scale $\lambda_{\rm D}$, which depends on the interaction processes which characterise the DM particle model.

For example, a DM-neutrino coupling would erase small DM structures due to collisional damping effects and late neutrino decoupling, see refs.~\cite{Boehm:2000gq, Boehm:2003hm, Boehm:2004th}. Simulations of non-linear structure formation in such models were analysed in ref.~\cite{Boehm:2014vja}. In refs.~\cite{Wilkinson:2013kia, Wilkinson:2014ksa}, the authors showed that a DM-photon or DM-neutrino coupling, with a cross section compatible with the CMB, could alleviate the \textit{missing satellite problem} \cite{Klypin:1999uc}. Moreover, effects related to velocity dispersion have proved to be relevant, e.g. in the Tseliakhovich-Hirata effect \cite{Tseliakhovich:2010bj}, where the relative motion among baryons and DM is considered. Se also refs.~\cite{Visbal:2012aw, Fialkov:2014rba, Hlozek:2014lca, Marsh:2015daa}. 



After kinetic decoupling, WIMPs ``free stream" but still possess a non-negligible velocity dispersion that fixes the free-streaming scale $\lambda_{\rm fs}$ below which the gravitational collapse is suppressed. In reference \cite{Green:2005fa} the authors derive analytically an approximate expression for the free-streaming scale, namely, $\lambda_{\rm fs} \sim 1$ pc and for the diffusion scale $\lambda_{\rm D} \sim 10^{-2}$ pc. A more detailed analysis can be found in \cite{Profumo:2006bv}. A different approach is adopted in ref.~\cite{Piattella:2013cma} where the authors consider first order fluctuations in the velocity dispersion, after kinetic decoupling. These may behave as pressure perturbations, allowing one to define a Jeans scale for WIMPs that is slightly larger than $\lambda_{\rm fs}$. Consequently, the cut-off of the matter fluctuation power spectrum at small scales is controlled by the Jeans scale deriving from velocity dispersion fluctuations.

If DM particles decouple relativistically from the cosmic plasma, they constitute the model dubbed Warm Dark Matter (WDM) \cite{Bardeen:1985tr, Viel:2005qj, Viel:2006kd, deVega:2013ysa, Destri:2013hha, deVega:2013jfy, Schneider:2013wwa, Dev:2013yza}. A thermal origin implies that particle masses are of the order of few eV in order to explain the observed abundance while a non-thermal origin allows masses of the order of few keV \cite{Giudice:1999am, Abazajian:2012ys, Okada:2012gf, Allahverdi:2012wb}. See also refs.~\cite{Berezhiani:1989fu, Berezhiani:1992rk, Khlopov:2004tn}.

WDM may alleviate some problems that plague CDM at small scales such as \textit{the missing satellite problem} \cite{Klypin:1999uc}, \textit{the core-cusp problem} \cite{deBlok:2009sp} or \textit{the too big to fail problem} \cite{BoylanKolchin:2011dk}. WDM effects related to the \textit{the core-cusp problem} were studied by high resolution simulations performed in ref.~\cite{Maccio2012}. These simulations have shown that the observed size of cores can be reproduced if the mass of WDM particles is about 0.1 keV. However such a low mass prevents the formation of dwarf galaxies since introduces a cut-off in the matter power spectrum at scales less than $10^{10}~M_\odot$.

On one hand, supporters of WDM, see for example refs.~\cite{deVega:2013ysa, Destri:2013hha, deVega:2013jfy}, argue that quantum effects on scales smaller that $100$ pc are relevant, and when properly taken into account, e.g. via a Thomas-Fermi approach, permit to predict sizes of galactic cores, masses of galaxies, velocity dispersions and density profiles in agreement with observations. On the other hand, recent investigations of WDM based on  the \textit{flatness} problem of the HI velocity-width function in the local universe and Lyman-$\alpha$ forest \cite{Schneider:2013wwa}, fail to alleviate small scale problems related to structure formation. According to \cite{Schneider:2013wwa}, this failure is probably due to the cut-off in the linear power spectrum that is too steep to simultaneously reproduce the Lyman-$\alpha$ data and to match dwarf galaxy properties. See also the very recent work \cite{Baur:2015jsy} for the tightest bound to date on WDM mass.

In this paper, the effects of velocity dispersion on the linear growth of DM fluctuations are analysed. We start from the Vlasov-Einstein equation for DM, building a hierarchy of equations that are truncated at the second order momentum, by neglecting contributions of the order of $\mathcal{O}(p^3/m^3)$, where $p$ and $m$ are respectively the proper momentum and the mass of a DM particle. Then, a modified CAMB code \cite{Lewis:1999bs} was used to compute the linear matter power spectrum. The derived results are compared with the matter power spectrum derived from a WDM model constituted by particles of mass $\sim$ 3.3 keV. Both models are able to produce a cut-off in scales of about $\sim$ 10h Mpc$^{-1}$ but for different physical reasons. We speculate if velocity dispersion effects would be more appropriate for solving small-scale issues present in the DM paradigm. In this case, a late non-thermal DM origin is required and our picture becomes similar to the Late-Forming scenario, in which DM particles are produced by the decay of the kinetic energy of a scalar field in a metastable vacuum.

The paper is organized as follows. In sec.~\ref{sec:2} the equations describing the evolution of small CDM fluctuations in the presence of a non-zero, small velocity dispersion are presented. In sec.~\ref{sec:3} the modifications of the DM power spectrum and the CMB TT power spectrum induced by the presence of velocity dispersion fluctuations are discussed. Finally, in sec~\ref{sec:4} the main results and conclusions are given. 
 
\section{Evolution equations for small dark matter fluctuations}\label{sec:2}

In this section we derive the equations governing the evolution of DM fluctuations in presence of a velocity dispersion.
We write the perturbed Friedmann-Lema\^itre-Robertson-Walker metric in the following way:
\begin{equation}
ds^2 = a^2(\tau)\left[-d\tau^2 + \left(\delta_{ij} + h_{ij}\right)dx^idx^j\right]\;,
\end{equation}
where $h_{ij}$, focusing only on scalar perturbations, is written in the following form as an inverse Fourier transform:
\begin{equation}
h_{ij}(\textbf{x},\tau) = \int d^3k\exp(i\textbf{k}\cdot\textbf{x})\left[\hat{k}_i\hat{k}_jh(\textbf{k},\tau) + \left(\hat{k}_i\hat{k}_j - \frac{1}{3}\delta_{ij}\right)6\eta(\textbf{k},\tau)\right]\;,
\end{equation}
where $\textbf{x}$ is the position 3-vector, $\textbf{k}$ is the wavenumber 3-vector with modulus $k$ and direction $\hat{k}_i$, and $h$ and $\eta$ are the scalar perturbations of the metric, i.e. the trace and traceless part, respectively. Note that, by definition, $\delta_{ij}\hat{k}^i\hat{k}^j = 1$.

We are going to determine the evolution equation for DM with velocity dispersion by taking up to the second momentum of the Boltzmann equation. Defining $f = f_0 + f_1$ as the one-particle distribution function, separated in its background unperturbed value and a small fluctuation about it, the perturbed Boltzmann equation can be computed, see ref.~\cite{Ma:1995ey}, as follows:
\begin{equation}\label{boltzeq}
\frac{\partial f_1}{\partial\tau} - \frac{\dot{a}}{a}p\frac{\partial f_1}{\partial p} + \frac{p}{E}\left(i\textbf{k}\cdot\hat{\textbf{n}}\right)f_1 + p\frac{\partial f_0}{\partial p}\left[\dot{\eta} - \frac{\dot{h} + 6\dot{\eta}}{2}(\hat{\textbf{k}}\cdot\hat{\textbf{n}})^2\right] = 0\;,
\end{equation}
where $p^i = p_i$ is the proper momentum; $\hat{n}^i$ is the unit vector describing the direction of $p^i$, i.e. $p^i = p\hat{n}^i$, where $p$ is the modulus of $p^i$, i.e. $p^2 = \delta_{ij}p^ip^j$ (remember that the definition of proper momentum incorporates the metric); $E^2 = p^2 + m^2$ is the energy; $\textbf{k}$ is the wave-vector coming from Fourier transformation and the dot denotes derivation with respect to the conformal time $\tau$. 


In the following, we consider momenta of Boltzmann equation~\eqref{boltzeq} in order to derive the  evolution equations for the perturbative quantities. This task is similar to the usual one for photons and neutrino. On the other hand, in our model for DM we will consider non-relativistic ``quasi-cold" particles, in the sense that they have a non-vanishing but small velocity dispersion. 

We implement this condition by employing the following expansion for the particle energy:
\begin{equation}
 E = m\left[1 + \frac{p^2}{2m^2} + O\left(p^4/m^4\right)\right]\;,
\end{equation}
and by neglecting terms of order $O\left(p^3/m^3\right)$. We will use the symbol $\sim$ when making this truncation. This will allow us to truncate the hierarchy of the Vlasov equation at the third order momentum. 


\subsection{Zero-order momentum of the perturbed Boltzmann equation}

Taking the zero-order momentum of eq.~\eqref{boltzeq} means to multiply it by $d^3p$ and to integrate in the proper momentum space. The result is:
\begin{equation}\label{zeromom}
 \frac{\partial n_1}{\partial\tau} + 3\frac{\dot a}{a}n_1 + ik_in_0v^i - 3n_0\dot{\eta} + \frac{\dot{h} + 6\dot{\eta}}{2}n_0 = 0\;,
\end{equation}
where we used the definitions
\begin{equation}\label{numdensdef}
n \equiv n_0 + n_1\;, \quad n_0 \equiv \int d^3p f_0\;, \quad n_1 \equiv \int d^3p f_1\;,
\end{equation}
for the particle number density\footnote{Note that the number of spin states, usually denoted by $g$, is incorporated in $f$.} and
\begin{equation}\label{bulkveleq}
n_0v^i \equiv \int d^3p \frac{p\hat{n}^i}{E}f_1\;,
\end{equation}
for the bulk velocity, which is a pure first-order quantity. We also used the geometrical property
\begin{equation}\label{propninj}
\int d\Omega \hat{n}^i\hat{n}^j = \frac{4\pi}{3}\delta^{ij}\;,
\end{equation}
in order to integrate by parts the last term in eq.~\eqref{boltzeq}. Defining the number density contrast as $\delta_n \equiv n_1/n_0$, we can rewrite eq.~\eqref{zeromom} as:
\begin{equation}\label{deltaneq}
\dot{\delta}_n + ik_iv^i + \frac{\dot{h}}{2} = 0\;,
\end{equation}
where we also used the background result $\dot{n}_0 + 3\dot{a}n_0/a = 0$, i.e. particle number conservation. For CDM, $\delta_n = \delta \equiv \rho_1/\rho_0$, but in our case, considering a non-vanishing velocity dispersion, that is no longer true. The energy density can be calculated as follows:
\begin{equation}
 \rho \equiv \int d^3p E f = \int d^3p \sqrt{p^2 + m^2} (f_0 + f_1) \sim m\int d^3p \left(1 + \frac{p^2}{2m^2}\right)(f_0 + f_1)\;.
\end{equation}
Using eq.~\eqref{numdensdef}, the above integrals can be rewritten as follows:
\begin{equation}\label{rhodef}
\rho = \rho_0 + \rho_1\;, \quad \rho_0 = n_0m + \frac{1}{2}n_0m\sigma^2_0\;, \quad \rho_1 = n_1m + \frac{1}{2}n_0mv^2_1\;,
\end{equation}
where $\sigma^2_0$ is the background velocity dispersion and $v^2_1$ its perturbative counterpart, defined as follows: 
\begin{equation}\label{v12def}
	n_0\sigma_0^2 \equiv \int d^3p\;\frac{p^2}{E^2}f_0 \sim \int d^3p\frac{p^2}{m^2}f_0\;, \quad
	n_0v_1^2 \equiv \int d^3p\frac{p^2}{E^2}f_1 \sim \int d^3p\frac{p^2}{m^2}f_1\;.
\end{equation}
From eq.~\eqref{rhodef} we can read out the density contrast:
\begin{equation}\label{deltadeltan}
\delta \equiv \frac{\rho_1}{\rho_0} =  \frac{n_1m + mn_0v_1^2/2}{n_0m + mn_0\sigma^2_0/2} = \frac{\delta_n + v^2_1/2}{1 + \sigma^2_0/2}\;,
\end{equation}
and rewrite eq.~\eqref{deltaneq} as follows:
\begin{equation}\label{deltaeq}
\left(1 + \frac{\sigma^2_0}{2}\right)\dot{\delta} - \frac{\dot a}{a}\sigma_0^2\delta - \frac{1}{2}\frac{d(v^2_1)}{d\tau} + ik_iv^i + \frac{\dot{h}}{2} = 0\;,
\end{equation}
where we used the result $\sigma^2_0 \propto a^{-2}$.

\subsection{First-order momentum of the perturbed Boltzmann equation}

In order to compute the first-order momentum of eq.~\eqref{boltzeq}, we multiply it by $d^3p p \hat{n}^i/E$ and integrate. This procedure gives the following result, using also eq.~\eqref{bulkveleq}:
\begin{equation}\label{mom1}
 \frac{\partial(n_0v^i)}{\partial\tau} + 4\frac{\dot{a}}{a}n_0v^i + ik_j\int d^3p\frac{p^2}{E^2}\hat{n}^i\hat{n}^jf_1 = 0\;,
\end{equation}
where we have used the fact that the integration of $f_0$ with an odd number of unit vectors $\hat{n}^i$ is vanishing because of the isotropy of the background solution. The last integral of eq.~\eqref{mom1} defines the perturbed velocity dispersion tensor:
\begin{equation}
 n_0({v^2_1})^{ij} \equiv \int d^3p\frac{p^2}{E^2}\hat{n}^i\hat{n}^jf_1 \sim \int d^3p\frac{p^2}{m^2}\hat{n}^i\hat{n}^jf_1\;,
\end{equation}
consistent with eq.~\eqref{v12def}, since $\delta_{ij}\hat{n}^i\hat{n}^j = 1$. Contracting eq.~\eqref{mom1} with $k_i$ and using the background solution for $n_0$, one obtains:
\begin{equation}\label{mom1bis}
 k_i\dot{v}^i + \frac{\dot{a}}{a}k_iv^i + ik_ik_j({v^2_1})^{ij} = 0\;.
\end{equation}

\subsection{Second-order momentum of the perturbed Boltzmann equation}

In order to compute the second-order momentum of eq.~\eqref{boltzeq}, we multiply it by $p^2\hat{n}^i\hat{n}^jd^3p/E^2$ and integrate. One finds:
\begin{equation}\label{mom2}
 \frac{\partial[n_0({v^2_1})^{ij}]}{\partial\tau} + 5\frac{\dot{a}}{a}n_0({v^2_1})^{ij} - 5\dot{\eta}n_0({\sigma^2_0})^{ij} + \frac{5(\dot{h} + 6\dot{\eta})}{2}\hat{k}_l\hat{k}_m\int d^3p\frac{p^2}{E^2}f_0\hat{n}^i\hat{n}^j\hat{n}^l\hat{n}^m = 0\;,
\end{equation}
where we used the definition
\begin{equation}
n_0({\sigma^2_0})^{ij} = \int d^3p \frac{p^2}{E^2}f_0\hat{n}^i\hat{n}^j \sim \int d^3p \frac{p^2}{m^2}f_0\hat{n}^i\hat{n}^j\;,
\end{equation}
for the background velocity dispersion tensor. In order to calculate the last integral of eq.~\eqref{mom2}, we need to compute first the following integral:
\begin{equation}
I^{ijlm} \equiv \int d\Omega \hat{n}^i\hat{n}^j\hat{n}^l\hat{n}^m\;.
\end{equation}
From eq.~\eqref{propninj} it is not difficult to determine that
\begin{equation}
I^{ijlm} = \frac{4\pi}{15}\left(\delta^{ij}\delta^{lm} + \delta^{il}\delta^{mj} + \delta^{im}\delta^{jl}\right)\;.
\end{equation}
So we can now cast eq.~\eqref{mom2} in the following form:
\begin{equation}\label{mom2bis}
 \frac{d({v^2_1})^{ij}}{d\tau} + 2\frac{\dot{a}}{a}({v^2_1})^{ij} - 5\dot{\eta}({\sigma^2_0})^{ij} + \frac{\dot{h} + 6\dot{\eta}}{6}\left(\delta^{ij} + 2\hat{k}^i\hat{k}^j\right)\sigma^2_0 = 0\;.
\end{equation}
The Boltzmann hierarchy is thus truncated and we have a finite number of equation (4 instead of the usual 2 for CDM) describing our ``quasi-cold" DM model. 

\subsection{Summary of the equations}

Considering scalar perturbations only for the velocity, i.e. $ik_iv^i = kv$, we can summarize the three evolution equations~\eqref{deltaneq}, \eqref{mom1bis} and \eqref{mom2bis} found so far:
\begin{eqnarray}
\label{deltaneq2} \dot{\delta}_n &+& kv + \frac{\dot{h}}{2} = 0\;,\\
\label{veq} \dot{v} &+& \frac{\dot{a}}{a}v - k\hat{k}_i\hat{k}_j({v^2_1})^{ij} = 0\;,\\
\label{v1eq} \frac{d({v^2_1}^{ij})}{d\tau} &+& 2\frac{\dot{a}}{a}({v^2_1})^{ij} - 5\dot{\eta}({\sigma^2_0})^{ij} + \frac{\dot{h} + 6\dot{\eta}}{6}\left(\delta^{ij} + 2\hat{k}^i\hat{k}^j\right)\sigma^2_0 = 0\;.
\end{eqnarray}
Now, we expand $({v^2_1})^{ij}$ in its trace and traceless part, i.e.
\begin{equation}
 ({v^2_1})^{ij} = \frac{v^2_1}{3}\delta^{ij} + \Sigma^{ij}\;,
\end{equation}
so that eq.~\eqref{veq} can be cast in the following form:
\begin{equation}
\dot{v} + \frac{\dot{a}}{a}v - \frac{k}{3}v_1^2 - k\Pi = 0\;,
\end{equation}
where we have defined the anisotropic shear as
\begin{equation}
\Pi \equiv \hat{k}_i\hat{k}_j\Sigma^{ij}\;.
\end{equation}
Moreover, we can separate eq.~\eqref{v1eq} in its trace and traceless contributions. The former is:
\begin{equation}
 \frac{d(v^2_1)}{d\tau} + 2\frac{\dot{a}}{a}v^2_1 + \frac{5\dot{h}}{6}\sigma^2_0 = 0\;,
\end{equation}
whereas the traceless part is:
\begin{equation}
\dot\Sigma^{ij} + 2\frac{\dot a}{a}\Sigma^{ij} + \frac{\dot h + 6\dot\eta}{3}\sigma^2_0\left(\hat k^i\hat k^j - \frac{1}{3}\delta^{ij}\right) = 0\;.
\end{equation}
Contracting the above equation with $\hat k_i\hat k_j$ we obtain an evolution equation for $\Pi$:
\begin{equation}
 \dot\Pi + 2\frac{\dot a}{a}\Pi + \frac{2}{3}\frac{\dot h + 6\dot\eta}{3}\sigma^2_0 = 0\;.
\end{equation}
We have thus a system of four equations for the six variables $\delta$, $v$, $v^2_1$, $\Pi$, $h$ and $\eta$:
\begin{eqnarray}
\label{deltaDmeqfin} \left(1 + \frac{\sigma^2_0}{2}\right)\dot{\delta} + \frac{\dot a}{a}\left(v^2_1 - \sigma^2_0\delta\right) + kv + \frac{\dot{h}}{2}\left(1 + \frac{5\sigma^2_0}{6}\right) = 0\;,\\
\label{veqfin} \dot{v} + \frac{\dot{a}}{a}v - \frac{k}{3}v^2_1 - k\Pi = 0\;,\\
\label{v1eqfin} \frac{d(v^2_1)}{d\tau} + 2\frac{\dot{a}}{a}v^2_1 + \frac{5\dot{h}}{6}\sigma^2_0 = 0\;,\\
\label{Pieqfin} \dot\Pi + 2\frac{\dot a}{a}\Pi + \frac{2}{3}\frac{\dot h + 6\dot\eta}{3}\sigma^2_0 = 0\;.
\end{eqnarray}
This system has to be coupled with the equations for the other components and with Einstein equations. Based on our assumptions of collision-less Boltzmann equation and truncation of the particle energy terms of order $\mathcal{O}(p^3/m^3)$, they describe the evolution of non-relativistic DM after kinetic decoupling. From Ref.~\cite{Ma:1995ey}, the Einstein equations can be cast in the following form:
\begin{eqnarray}
\label{k2etaeq} k^2\eta - \frac{1}{2}\frac{\dot a}{a}\dot h = 4\pi G a^2\delta T^0{}_0\;,\\
\label{k2dotetaeq} k^2\dot\eta = 4\pi G a^2ik^i\delta T^0{}_i\;,\\
\label{ddotheinsteq} \ddot h + 2\frac{\dot a}{a}\dot h - 2k^2\eta = -8\pi G a^2\delta T^i{}_{i}\;,\\
 \ddot h + 6\ddot\eta + 2\frac{\dot a}{a}(\dot h + 6\dot\eta) - 2k^2\eta = 24\pi G a^2\hat k_i\hat k^j(\delta T^i{}_j - \delta T^l{}_l\delta^i{}_j/3)\;,
\end{eqnarray}
where the perturbed stress energy tensor introduced above is the total one. Only the DM part is changed in our treatment, in the following way: 
\begin{eqnarray}
\delta T^0{}_0(\textrm{DM}) &=& - \int d^3pEf_1 = -\rho_1\;,\\
\delta T^0{}_i(\textrm{DM}) &=& \int d^3p p\hat{n}_if_1 \sim  n_0mv_i\;,\\
\delta T^i{}_j(\textrm{DM}) &=& \int d^3p \frac{p^2\hat{n}^i\hat{n}_j}{E}f_1 \sim n_0m({v^2_1})^{i}{}_j\;.
\end{eqnarray}
Therefore, the perturbed DM stress energy tensor combinations which enter the Einstein equations on the right hand sides are:
\begin{eqnarray}
 \delta T^0{}_0(\textrm{DM}) &=& -\rho_0\delta\;,\\
 ik^i\delta T^0{}_i(\textrm{DM}) &=& \rho_0\frac{kv}{1 + \sigma^2_0/2}\;,\\
 \delta T^i{}_i(\textrm{DM}) &=& \rho_0\frac{v^2_1}{1 + \sigma^2_0/2}\;,\\
 \hat k_i\hat k^j(\delta T^i{}_j - \delta T^l{}_l\delta^i{}_j/3)(\textrm{DM}) &=& \rho_0\frac{\Pi}{1 + \sigma^2_0/2}\;.
\end{eqnarray}
Combining eqs.~\eqref{deltaneq2}, \eqref{veqfin} and \eqref{v1eqfin} it is possible to find the following second-order equation for $\delta_n$:
\begin{equation}\label{2ndordereqdeltan}
	\ddot\delta_n + \frac{\dot a}{a}\dot\delta_n + k^2\left(\frac{v_1^2}{3} + \Pi\right) + \frac{\ddot{h}}{2} + \frac{\dot{a}}{a}\frac{\dot{h}}{2} = 0\;.
\end{equation}
The last two terms on the lhs are the usual gravitational potential contribution, which from eqs.~\eqref{k2etaeq} and \eqref{ddotheinsteq} reads:
\begin{equation}
	\frac{\ddot{h}}{2} + \frac{\dot{a}}{a}\frac{\dot{h}}{2} = 4\pi Ga^2\left(\delta T^0{}_0 - \delta T^i{}_i\right)\;.
\end{equation}
Note the term multiplying $k^2$, in eq.~\eqref{2ndordereqdeltan}. In general, it can be cast as a sum of a contribution dependent from $\delta_n$, thereby acting as an adiabatic pressure perturbation, plus an entropy contribution. Both affect the evolution of $\delta_n$ on small scales in a manner we show in the next section. The evolution equation for $\delta$ can be found from eq.~\eqref{2ndordereqdeltan} using the relation~\eqref{deltadeltan}.

\subsection{Fluid interpretation}

In Ref.~\cite{Ma:1995ey} the following two equations, coming from the stress-energy conservation equation can be found:
\begin{eqnarray}
 \dot\delta &=& -(1 + w)\left(\theta + \frac{\dot h}{2}\right) - 3\frac{\dot a}{a}\left(\frac{\delta P}{\delta\rho} - w\right)\delta\;,\\
 \dot\theta &=& -\frac{\dot a}{a}(1 - 3w)\theta - \frac{\dot w}{1 + w}\theta + \frac{\delta P/\delta\rho}{1 + w}k^2\delta - k^2\sigma\;.
\end{eqnarray}
These equations are equivalent to the first two momenta of the Boltzmann equation for DM that we have considered previously, viz. eqs.~\eqref{deltaDmeqfin} and \eqref{veqfin}, provided the following identifications are made: $kv = (1 + 5\sigma^2_0/6)\theta$ and $\Pi = -(1 + 5\sigma^2_0/6)\sigma$. The equation of state parameter and the pressure perturbation are given as follows as functions of $\sigma^2_0$ and $v^2_1$:
\begin{equation}\label{wdeltapeq}
 w \equiv \frac{P_0}{\rho_0} = \frac{\sigma^2_0/3}{1 + \sigma^2_0/2}\;,\qquad \frac{\delta P}{\rho_0} = \frac{v^2_1/3}{1 + \sigma^2_0/2}\;.
 \end{equation}
 Notice how velocity dispersion endows the fluid with a time-dependent equation of state and how $v^2_1$ plays the role of a pressure perturbation. The effective speed of sound $c^2_{\rm eff}\equiv P_1/\rho_1$ and the adiabatic speed of sound $c_a^2 \equiv \dot P_0/\dot\rho_0$ have the following forms:
\begin{equation}\label{sosvd}
 c_{\rm eff}^2 = \frac{v_1^2/3}{\delta_n + v_1^2/2}\;, \qquad c_a^2 = \frac{5\sigma_0^2/3}{3 + 5\sigma_0^2/2}\;.
\end{equation}
When perturbations are adiabatic, then $c_{\rm eff}^2 = c_a^2$, which gives: 
\begin{equation}\label{v1deltarelad}
v_1^2 = \frac{5}{3}\sigma_0^2\delta_n = \frac{5}{3}\sigma_0^2\frac{1 + \sigma_0^2/2}{1 + 5\sigma_0^2/6}\delta\;,
\end{equation}
where we used eq.~\eqref{deltadeltan}. With this condition, eq.~\eqref{v1eqfin} becomes:
\begin{equation}
\label{v1eqfinad} \dot\delta_n + \frac{\dot{h}}{2} = 0\;,
\end{equation}
which, when compared to eq.~\eqref{deltaneq}, implies $ik^iv_i = kv = 0$. In turn, this implies from eqs.~\eqref{veqfin} and \eqref{Pieqfin} that
\begin{eqnarray}
	\frac{v_1^2}{3} + \Pi = 0\;,\\
	\sigma_0^2(3\dot h + 8\dot\eta) = 0\;.
\end{eqnarray}
The last equation implies either a vanishing velocity dispersion, thereby frustrating our purpose, or it gives an unnatural relation between the geometric perturbations, i.e. $3\dot h = -8\dot\eta$. Therefore, the conclusion is that the perturbations cannot be adiabatic. We investigate this issue in some detail in the next subsection.
  
\subsection{Adiabaticity}

First of all, we write down the first law of thermodynamics:
\begin{equation}
TdS = dU + PdV\;,
\end{equation}
 where the variables have the usual meaning. Now consider $U = \rho V$ and $V = N/n$, where $N$ is the particle number, which we consider fixed i.e. no DM particle are being created (or annihilated). With these positions we can write:
 \begin{equation}
Tds = \frac{d\rho}{n} - \frac{\rho + P}{n^2}dn\;,
\end{equation}
where $s \equiv S/N$ is the entropy per particle. Remember that we assumed conserved particle number, therefore $N$ could enter in the differential $dS$. The density, pressure and number density which appear in the above equation are the total ones. If we ask adiabaticity for each single component, one gets for DM:
\begin{equation}
 \frac{\delta}{1 + P/\rho} = \delta_n\;,
\end{equation}
and using eq.~\eqref{wdeltapeq} one obtains again eqs.~\eqref{deltadeltan} and \eqref{v1deltarelad}, along with the related problems we already mentioned.

If we demand that the total entropy fluctuation is vanishing, we get that:
\begin{equation}
 \frac{\delta}{1 + w} = \frac{3}{4}\delta_\gamma\;,
\end{equation}
and similar equations for the other components (neutrinos and baryons). Now, on the background, energy density and particle number conservation ensures that $T_0\dot s_0 = 0$, i.e. adiabaticity. Note that $T_0/m = \sigma^2_0/3$.  At the first perturbative order, using also the equations for the perturbations that we derived above, one can find that
\begin{equation}
\frac{T_0}{m}\dot s_1 = \frac{5\sigma^2_0}{6}kv\;.
\end{equation}
Therefore, the initial condition $v = 0$ also implies that the initial time-derivative of DM entropy is vanishing.

\subsection{Initial conditions}

In order to properly modify CAMB we need not only to include the new equations containing the velocity dispersion, but also to treat properly the initial conditions (IC). From the discussion of the previous subsections, in our model we cannot have adiabatic IC.

Following e.g. \cite{Ma:1995ey} and \cite{Bucher:1999re} it is not difficult to see that the IC for $h$ and $\eta$ are not modified by the presence of a velocity dispersion since we are deep in the era where photons and neutrinos dominate and in our model we leave these components untouched. Therefore, $\delta_\gamma = \delta_\nu = -2h/3$.

Using eqs.~\eqref{deltaDmeqfin}-\eqref{Pieqfin} in the limit $k\tau \to 0$, we obtain the following IC for DM with velocity dispersion:
\begin{eqnarray}
\label{deltaIC} \left(1 - \frac{\sigma^2_0}{2}\right)\delta = -v^2_1 - \frac{h}{2}\left(1 + \frac{5\sigma^2_0}{6}\right)\;,\\
\label{vIC} v = 0\;,\\
\label{v1eqfinIC} v^2_1 = -\frac{5h}{18}\sigma^2_0\;,\\
\label{PieqfinIC} \Pi = -\frac{2}{9}\frac{h + 6\eta}{3}\sigma^2_0\;.
\end{eqnarray}
Of course, for a vanishing $\sigma_0^2$ one recovers the usual adiabatic IC.

\section{Numerical results}\label{sec:3}

We use as free parameter of the modified equations in CAMB the velocity dispersion of DM at the present moment, which we dub $\beta^2$. Since the background velocity dispersion scales as
\begin{equation}\label{sigma0evol}
\sigma_0^2 = \frac{\sigma_{0({\rm kd})}^2a_{\rm kd}^2}{a^2}\;,
\end{equation}
where $\sigma_{0({\rm kd})}^2$ and $a_{\rm kd}$ are respectively the velocity dispersion and the scale factor at the kinetic decoupling, $\beta^2$ is defined as
\begin{equation}\label{beta2}
\beta^2 = \sigma_{0({\rm kd})}^2a_{\rm kd}^2\;.
\end{equation}
If DM has been thermally produced, then our approximation of non-relativistic particles allows us to use a Maxwell-Boltzmann distribution for DM particles and to write $\sigma_{0({\rm kd})}^2 = 3T_{\rm kd}/m$, where $T_{\rm kd}$ is the cosmic plasma temperature at kinetic decoupling and $m$ is the particle mass. In this case we have for $\beta^2$: 
\begin{equation}
\beta^2 = \frac{3T_{\rm kd}}{m}\left(\frac{T_0}{T_{\rm kd}}\right)^2\left(\frac{g_0}{g_{\rm kd}}\right)^{2/3}\;,
\end{equation}
where $T_0$ is the present time CMB temperature and $g_0$ and $g_{\rm kd}$ are the effective degrees of freedom of the cosmic plasma. Using $T_0 = 2.725\mbox{ K} = 2.35\times 10^{-13}$ GeV, we have
\begin{equation}\label{eqconstr}
\beta^2 = 8.28\times 10^{-26}\left(\frac{g_0}{g_{\rm kd}}\right)^{2/3}\left(\frac{100 \mbox{ GeV}}{m}\right)\left(\frac{20 \mbox{ MeV}}{T_{\rm kd}}\right)\;,
\end{equation}
where we put in evidence the relevant numbers for the DM particle candidate, the neutralino. Note that the effective degrees of freedom $g_0$ and $g_{\rm kd}$ take into account the entropy conservation  when the cosmic
plasma is heated by annihilation of particles since kinetic decoupling.
Equation~\eqref{eqconstr} shows that if a constraint is imposed on $\beta^2$ this is equivalent to say that we are imposing 
a constraint on the product $m T_{\rm kd}$, which represents a ``finger-print" of the early interactions of DM particles with
the cosmic plasma.

\begin{figure}[htbp]
\centering
\includegraphics[width=0.7\columnwidth]{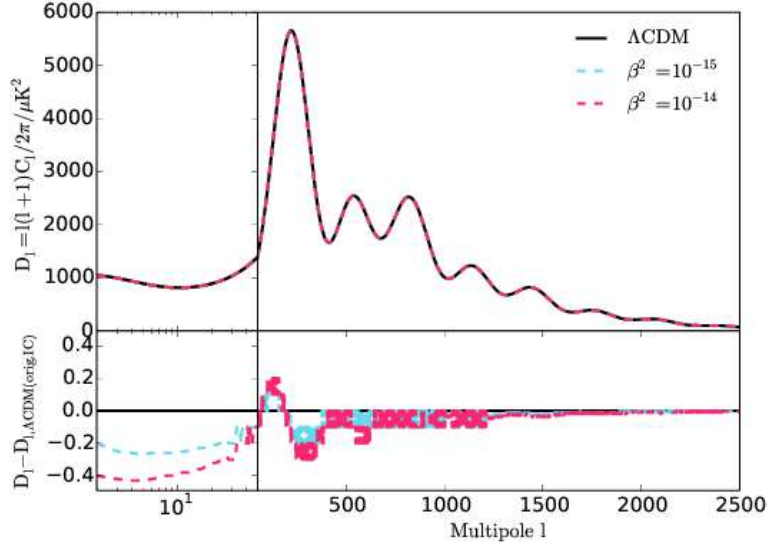}
\caption{Upper panel. The CMB TT power spectrum computed for $\beta^2 = 10^{-14}$ and $\beta^2 = 10^{-15}$. Lower panel. Residuals with respect to the reference $\Lambda$CDM model, represented with a dotted line.}
\label{Fig:CMB}
\end{figure}

\begin{figure}[htbp]
\centering
	\includegraphics[width=0.7\columnwidth]{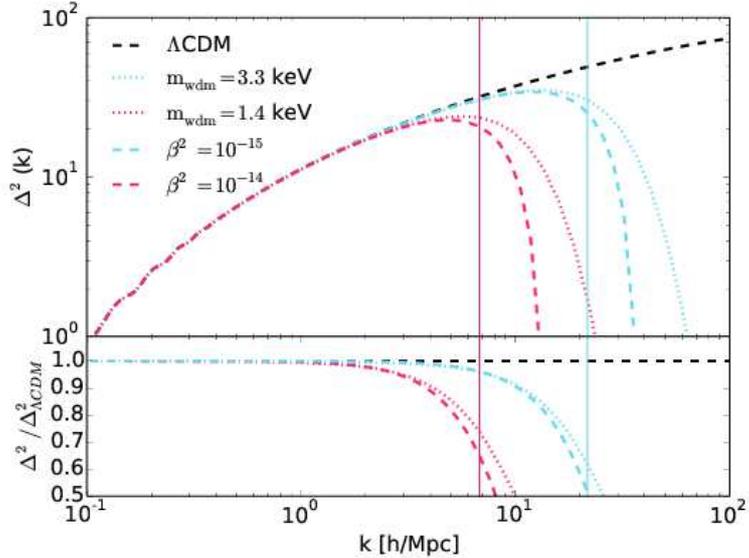}
	\caption{The DM power spectra at $z = 0$ computed for $\beta^2 = 10^{-15}$ (cyan dashed line) and $\beta^2 = 10^{-14}$ (magenta dashed line) compared with the corresponding ones for WDM particles of mass $3.3$ keV (cyan dotted line) and $1.4$ keV (magenta dotted line). For the latter we used the fitting formula of \cite{Viel:2005qj}. The vertical magenta and cyan solid lines determine the limiting scales, defined as those for which $\beta^2 = a^2$ at horizon crossing.}
\label{Fig:Matterpowerz0}
\end{figure}

We plot respectively in figs.~\ref{Fig:CMB} and \ref{Fig:Matterpowerz0} the CMB TT power spectrum and the DM power spectrum for $\beta^2 = 10^{-14}$ and $\beta^2 = 10^{-15}$ at $z = 0$. The standard CDM case ($\beta^2 = 0$) is also displayed for reference.
In the lower panel of fig.~\ref{Fig:CMB} residuals with respect to our adopted reference (the standard $\Lambda$CDM model) are shown. With only a tiny correction (0.4 over 1000, for the $\beta^2 = 10^{-14}$) the CMB spectrum does not effectively constrain the velocity dispersion of DM particles.
In fig.~\ref{Fig:Matterpowerz0} we compare the DM power spectra calculated for $\beta^2 = 10^{-15}$ and $\beta^2 = 10^{-14}$ with those derived from WDM particles with masses $m_{\rm wdm} = 1.4$ keV and $m_{\rm wdm} = 3.3$ keV. For the plots of the WDM power spectrum we used the fitting formula provided by \cite{Viel:2005qj}. The vertical solid magenta and cyan lines (for $\beta^2 = 10^{-14}$ and $\beta^2 = 10^{-15}$, respectively) represent the scales $k = 21.7$ and $k = 68.5$ h Mpc$^{-1}$, respectively, for which $\beta^2 = a^2$ at horizon crossing, thus constituting the limit of validity of our equations.

The masses of WDM particles were chosen from Lyman-$\alpha$ forest constraints \cite{Viel:2005qj, Viel:2006kd}. Then we computed the values for $\beta^2$ which produce similar cut-offs. The plots in fig.~\ref{Fig:Matterpowerz0} indicate that velocity dispersion effects in CDM may mimic the cut-off due to free-streaming, typical of WDM particles. \textit{En passant}, it is worth noting that, albeit such a cut-off is still consistent with constraints based on Lyman-$\alpha$ forest data, for masses larger than $3.3$ keV, WDM appears no better than CDM in solving the small scale CDM anomalies \cite{Schneider:2013wwa}. 

It is important to emphasize that although in the considered case WDM and CDM produce similar cut-offs in the linear power spectrum, different mechanisms are involved: in the case of WDM an exponential cut-off is produced in the distribution function due to free-streaming \cite{Boyanovsky:2010pw, Boyanovsky:2010sv} while velocity dispersion effects of CDM particles act as a pressure perturbation \cite{Piattella:2013cma}.

In order to mimic the cut-off in the linear power spectrum produced by 3.3 keV WDM particles it is required that $\beta^2 = 10^{-15}$. This condition cannot be satisfied by non-relativistic DM thermally produced. 

Therefore, let's consider the possibility that DM particles were produced non-thermally. In this case, $\beta^2$ fixes the instant when these particles become non-relativistic. Indeed, if we assume that the transition to the non-relativistic regime occurs when $\sigma_0^2 \sim 0.1$, then $\beta^2 = 10^{-15}$ implies that the transition happens when $z_c \sim 10^7$. 

Although this requires some fine-tuning in the production mechanism in order that primordial nucleosynthesis be not perturbed, this possibility may offer an issue for the problem of the excess of satellites. 

If DM has a non-thermal origin able to explain, for instance, a parameter $\beta^2 = 10^{-15}$, we have seen that the non-relativistic phase should begin around $z_c \sim 10^7$. This is similar to the \textit{Late Forming Dark Matter} scenario (LFDM) proposed in ref.~\cite{Das:2006ht}, see also \cite{Sarkar:2014bca}. In the LFDM picture a scalar field trapped in a metastable vacuum begins to oscillate and suffers a transition in which its equation of state passes from ``radiation'' ($w \sim$ 1/3) to ``cold matter'' ($w \sim$ 0).  The matter power spectrum in this model typically shows a sharp break at small scales, below which power is suppressed \cite{Das:2006ht, Agarwal:2014qca}. Cosmological simulations adopting this scenario were performed in \cite{Agarwal:2014qca}, who have assumed that the transition occurred at $z_t = 1.5\times 10^6$, comparable to the redshift required to have $\beta^2 = 10^{-15}$. Comparing our fig.~\ref{Fig:Matterpowerz0} with fig.~1 of reference \cite{Agarwal:2014qca}, in which is shown the matter power spectrum resulting from the LFDM model as well as those resulting from WDM and $\Lambda$CDM models, it is interesting to notice that the present model, including velocity dispersion effects, produces a power spectrum similar to the LFDM picture. 

\begin{figure}[htbp]
\centering
	\includegraphics[width=0.7\columnwidth]{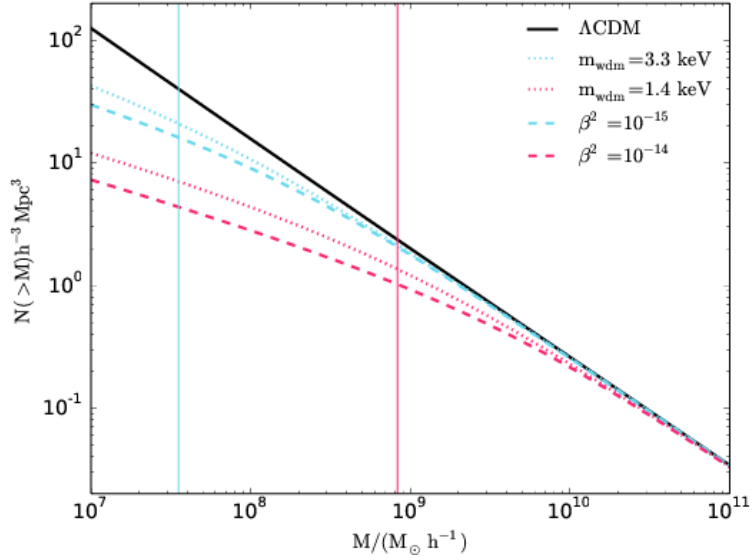}
	\caption{Number count as function of the mass scale computed for $\beta^2 = 10^{-15}$ (cyan dashed line) and $\beta^2 = 10^{-14}$ (magenta dashed line) compared with the corresponding ones for WDM particles of mass $3.3$ keV (cyan dotted line) and $1.4$ keV (magenta dotted line).}
	\label{Fig:massf}
\end{figure}

\begin{figure}[htbp]
\centering
	\includegraphics[width=0.7\columnwidth]{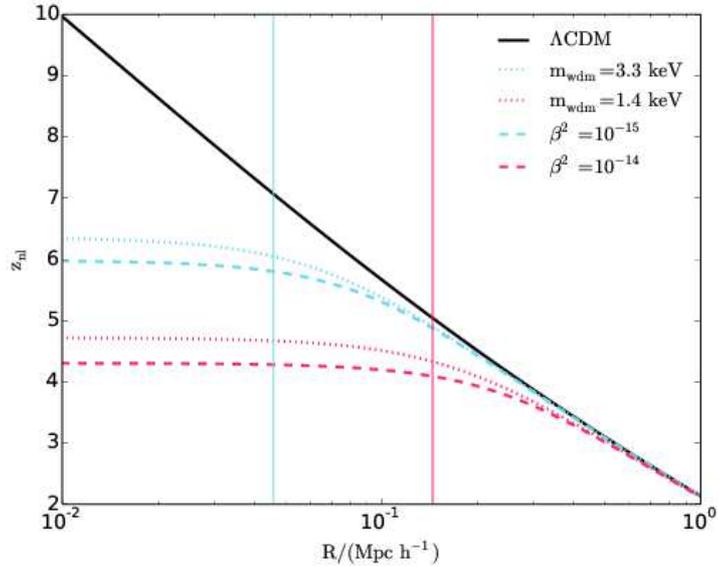}
	\caption{Redshift at which the mass-variance attains unity as function of the scale of the fluctuation computed for $\beta^2 = 10^{-15}$ (cyan dashed line) and $\beta^2 = 10^{-14}$ (magenta dashed line) compared with the corresponding ones for WDM particles of mass $3.3$ keV (cyan dotted line) and $1.4$ keV (magenta dotted line).}
	\label{Fig:znrl}
\end{figure}

In fig.~\ref{Fig:massf} is plotted the density of halos above a given mass scale $M$ as a function of the mass for $z = 0$. The integral mass spectrum was computed for $\beta^2 = 0$ ($\Lambda$CDM model), $\beta^2 = 10^{-15}$ and
$\beta^2 = 10^{-14}$. For comparison, it is also plotted the integral mass spectrum for WDM models having particles with masses respectively equal to 1.4 keV and 3.3 keV. Note that for the case $\beta^2 = 10^{-15}$ there is a slight deficiency of halos for masses less than $2\times 10^9~M_\odot$, comparable to the WDM model with 3.3 keV particles. In fig.~\ref{Fig:znrl} it is shown, for the same aforementioned parameters, the redshift at which the mass variance is equal to the unity as functions of the scale. A top-hat filter normalized to the Planck 2015 result, that is $\sigma_8 = 0.829 \pm 0.014$ \cite{Ade:2015xua} was used in the computations. 

\subsection{The mass of dark matter particles in the case $\beta = 10^{-15}$}

We have seen that $\beta^2 = 10^{-15}$ can be explained only if dark matter particles have a non-thermal origin otherwise the predicted cosmic abundance would be inconsistent with the observed one. In this case, the fiducial value that fixes the initial conditions for the dispersion velocity corresponds to the transition between relativistic and non-relativistic regimes or corresponds to a ``late formation of (non-relativistic) dark matter particles". In both cases, the fiducial value corresponds to a critical redshift of about $z_c \approx 10^7$ or to a temperature of about $2.4$ keV, thus after the primordial nucleosynthesis.

What is the mass of particles related to those possible scenarios? Just after the critical redshift the distribution function obeys the Einstein-Vlasov equation whose solution for a flat FRW model is of the form $f = f(ap/p_0)$ with $ap/p_0 =$ constant. The constant $p_0$ has the dimension of a momentum and was introduced in order that the argument of the distribution function be dimensionless. In this case, the density of dark matter particles is given by
\begin{equation}
	n = \frac{1}{2\pi^2}\int_0^\infty dp\; p^2 f(ap/p_0) = \frac{p_0^3}{2\pi^2a^3}I_2\;,
\end{equation}
where we introduced the integral function
\begin{equation}
	I_k = \int_0^\infty dx\; x^k f(x)\;.
\end{equation}
The dimensionless velocity dispersion, in the non-relativistic case, is given by
\begin{equation}
	\sigma^2 = \frac{1}{2\pi^2n}\int_0^\infty dp\;p^2\frac{p^2}{m^2}f(ap/p_0) = \frac{p_0^2}{m^2a^2}\frac{I_4}{I_2}\;.
\end{equation}
Thus, the parameter $\beta^2$ is simply
\begin{equation}
	\beta^2 = \sigma^2a^2 = \frac{p_0^2}{m^2}\frac{I_4}{I_2}\;.
\end{equation}
In order to estimate $p_0$ we require that these particles have the adequate cosmic abundance, in other words they have to satisfy
\begin{equation}
	\Omega_x = \frac{8\pi G}{3H_0^2}n_0m = \frac{4}{3\pi}\frac{Gmp_0^3}{H_0^2}I_2\;.
\end{equation}
Eliminating $p_0$, one obtains:
\begin{equation}
	\beta^2 = 3.45 \times 10^{-15}\frac{I_4}{I_2^{5/3}}\left(\frac{\Omega_x h^2}{0.12}\right)^{2/3}\left(\frac{\mbox{keV}}{m}\right)^{8/3}\;.
\end{equation}
This result is general, being the details of the particle distribution function encoded in the ratio $I_4/I_2^{5/3}$, which
involves integrals depending on the distribution function.

\section{Conclusions}\label{sec:4}

In this paper we report an study on velocity dispersion effects on the evolution of small perturbations of DM density. We considered the perturbed Vlasov-Einstein equation for DM an computed its momenta up to the second one, truncating the hierarchy based on the ``quasi-cold" condition, for which we neglect terms $\mathcal{O}(p^3/m^3)$. We implement this set of equations in CAMB thereby computing the CMB TT power spectrum as well as the matter power spectrum. 

For DM candidates issued from extensions of the standard model like the neutralino, no significant velocity dispersion effects are expected either in the CMB TT power spectrum or in the matter power spectrum. Interactions with leptons keep these particles coupled thermally until temperatures of the order of 10-20 MeV. Consequently the expected value of the parameter $\beta^2$ is around $10^{-25}$, several orders of magnitude smaller than the values required to affect the CMB TT power spectrum or the linear power spectrum.

However, if DM particles are produce lately ($z_{\rm c} \sim 10^7$) and non-thermally, higher values of the parameter $\beta^2$ are possible. In this case, velocity dispersion effects introduce a cut-off in the linear matter power spectrum at small scales, similar to that produced by WDM having particles of mass around 3.3 keV. It is important to emphasize that the physical nature of the damping effects due to WDM and velocity dispersion of CDM is not the same: the former is related to ``free-streaming" while the latter mimics pressure effects, see ref.~\cite{Piattella:2013cma} for a discussion on this point. The present model is comparable to the Late Forming dark matter scenario, in which non- relativistic matter appear as consequence of the decay of a scalar field.

%

\acknowledgments
JAFP thanks the program {\it Science without Borders} of the Brazilian Government for the financial support to this research and the Federal University of the Esp\'irito Santo State (Brazil) for his hospitality. The authors thank the anonymous referee for a thorough analysis of our manuscript which led to remarkable improvements. The authors thank CNPq and FAPES for support.

\bibliographystyle{JHEP}
\bibliography{Vlasovbib}

\end{document}